\documentclass[
reprint,           
superscriptaddress,
amsmath,           
amssymb,           
aps,               
prl,               
notitlepage,       
longbibliography,  
floatfix,          
nofootinbib,
]{revtex4-1}

\usepackage{tensor}     
\usepackage{float}
\usepackage[caption = false]{subfig} 
\usepackage[final]{graphicx}   
\usepackage[
colorlinks=true,        
citecolor=blue,         
linkcolor=blue,         
urlcolor=blue           
]{hyperref}             
\usepackage{bm}         
\usepackage{xcolor}     
\usepackage{lipsum}
\usepackage{color}      
\usepackage[utf8]{inputenc} 
\usepackage[section]{placeins} 

\newcommand{\nc}{\newcommand*} 

\nc{\al}{\alpha}
\nc{\s}{\sigma}
\nc{\dt}{\delta}
\nc{\Dt}{\Delta}
\nc{\Ld}{\Lambda}
\nc{\p}{\partial}
\nc{\om}{\omega}
\nc{\Om}{\Omega}
\nc{\rd}{\mathrm{d}}
\nc{\Od}[1]{\mathcal{O}(#1)} 
\nc{\kp}{\kappa}

\def\({\left(}
\def\){\right)}
\def\[{\left[}
\def\]{\right]}
\def\e{\begin{equation}}
\def\q{\end{equation}}
\def\m{\begin{eqnarray}}
\def\n{\end{eqnarray}}
\nc{\Eq}[1]{Eq.~\eqref{#1}}     
\nc{\Fig}[1]{Fig.~\ref{#1}}     
\nc{\Table}[1]{Table~\ref{#1}}  
\nc{\Sec}[1]{Sec.~\ref{#1}}     
\nc{\Msun}{M_\odot}             
\nc{\fpbh}{f_{\mathrm{pbh}}}    
\nc{\fpbhn}{f_{\mathrm{pbh0}}}    
\nc{\mR}{\mathcal{R}} 
\nc{\seq}{\sigma_{\mathrm{eq}}}
\nc{\ogw}{\Omega_{\mathrm{GW}}}
\nc{\gpcyr}{\mathrm{Gpc}^{-3}\,\mathrm{yr}^{-1}}
\nc{\lvc}{LIGO/Virgo} 
\nc{\SNR}{\mathrm{SNR}} 
\nc{\mmin}{{m_{\mathrm{min}}}}
\nc{\mmax}{{m_{\mathrm{max}}}}
\nc{\Mmin}{{M_{\mathrm{min}}}}
\nc{\fmin}{{f_{\mathrm{min}}}}
\nc{\VT}{\mathrm{VT}}
\nc{\rhoGW}{\rho_{\mathrm{GW}}}
\nc{\vth}{\vec{\theta}}
\nc{\vd}{\vec{d}}
\nc{\vla}{\vec{\lambda}}
\nc{\Nobs}{N_{\mathrm{obs}}}
\nc{\av}[1]{\langle #1 \rangle} 
\nc{\km}{\mathrm{km}}
\nc{\Mpc}{\mathrm{Mpc}}
\nc{\Tobs}{T_{\mathrm{obs}}}
\nc{\Ntemp}{N_{\mathrm{temp}}}
\nc{\addref}{[\textcolor{red}{add ref}] } 
\nc{\eg}{\textit{e.g.~}}
\nc{\app}{\approx}
\nc{\hf}{\frac{1}{2}}
\nc{\discuss}{\textcolor{red}{Add discussion here!}}
\nc{\red}[1]{\textcolor{red}{#1}}
\nc{\mH}{\mathcal{H}}
\nc{\cs}{c_s^2}
\nc{\Sij}[1]{S_{ij}^{(#1)}}
\nc{\vi}[1]{v_i^{(#1)}}
\nc{\no}{\nonumber}
\def\<{\left\langle}
\def\>{\right\rangle}

\nc{\bk}{\bm{k}}
\nc{\bq}{\bm{q}}
\nc{\bp}{\bm{p}}
\nc{\bl}{\bm{l}}
\nc{\bx}{\bm{x}}
\nc{\be}{\mathbf{e}}
\nc{\mS}{\mathcal{S}}
\nc{\te}{\tilde{\eta}}
\nc{\tp}{\tilde{p}}
\nc{\tk}{\tilde{k}}
\nc{\tx}{\tilde{x}}
\nc{\tF}{\tilde{F}}
\nc{\tA}{\tilde{A}}
\nc{\mkpq}{|\bk-\bp-\bq|}
\nc{\mpq}{|\bp-\bq|}
\nc{\mkp}{|\bk-\bp|}
\nc{\mSi}[1]{\mS^{(#1)}({\bk, \eta})}
\nc{\vk}{\vec{k}}
\nc{\kstar}{k_*}
\nc{\fstar}{f_*}
\nc{\xstar}{x_*}
\nc{\mpbh}{m_{\rm{pbh}}}
\nc{\bn}[1]{\dt\bm{t}_{\text{#1}}}
\nc{\bC}[1]{\bm{C}_{\text{#1}}}
\nc{\NTOA}{N_{\text{TOA}}}
\nc{\Nmode}{{N_{\text{mode}}}}
\nc{\ARN}{A_{\rm{RN}}}
\nc{\gRN}{\gamma_{\rm{RN}}}
\nc{\bS}{\mathbf{\Sigma}}
\nc{\br}{\mathbf{r}}
\nc{\bN}{\mathbf{R}}
\nc{\arXiv}[2]{\href{http://arxiv.org/pdf/#1}{{\tt [#2/#1]}}}
\nc{\arXivold}[1]{\href{http://arxiv.org/pdf/#1}{{\tt [#1]}}}

\renewcommand{\vec}[1]{\boldsymbol{#1}} 

\begin{document}
	
\title{Pulsar Timing Array Constraints on Primordial Black Holes with NANOGrav 11-Year Data Set}
	
\author{Zu-Cheng Chen}
\email{chenzucheng@itp.ac.cn} 
\affiliation{CAS Key Laboratory of Theoretical Physics, 
	Institute of Theoretical Physics, Chinese Academy of Sciences,
	Beijing 100190, China}
\affiliation{School of Physical Sciences, 
	University of Chinese Academy of Sciences, 
	No. 19A Yuquan Road, Beijing 100049, China}
\affiliation{School of Physics and Astronomy, Cardiff University, 
    Cardiff CF24 3AA, United Kingdom}
\author{Chen Yuan}
\email{yuanchen@itp.ac.cn}
\affiliation{CAS Key Laboratory of Theoretical Physics, 
	Institute of Theoretical Physics, Chinese Academy of Sciences,
	Beijing 100190, China}
\affiliation{School of Physical Sciences, 
	University of Chinese Academy of Sciences, 
	No. 19A Yuquan Road, Beijing 100049, China}

\author{Qing-Guo Huang}
\email{huangqg@itp.ac.cn}
\affiliation{CAS Key Laboratory of Theoretical Physics, 
	Institute of Theoretical Physics, Chinese Academy of Sciences,
	Beijing 100190, China}
\affiliation{School of Physical Sciences, 
	University of Chinese Academy of Sciences, 
	No. 19A Yuquan Road, Beijing 100049, China}
\affiliation{School of Fundamental Physics and Mathematical Sciences
Hangzhou Institute for Advanced Study, UCAS, Hangzhou 310024, China}
\affiliation{Center for Gravitation and Cosmology, 
	College of Physical Science and Technology, 
	Yangzhou University, Yangzhou 225009, China}
\affiliation{Synergetic Innovation Center for Quantum Effects and Applications, 
	Hunan Normal University, Changsha 410081, China}
	
\date{\today}

\begin{abstract}
The detection of binary black hole coalescences by LIGO/Virgo has aroused the interest in primordial black holes (PBHs), because they could be both the progenitors of these black holes and a compelling candidate of dark matter (DM). PBHs are formed soon after the enhanced scalar perturbations re-enter horizon during radiation dominated era, which would inevitably induce  gravitational waves as well. Searching for such scalar induced gravitational waves (SIGWs) provides an elegant way to probe PBHs. We perform the first direct search for the signals of SIGWs accompanying the formation of PBHs in North American Nanohertz Observatory for Gravitational waves (NANOGrav) 11-year data set. No statistically significant detection has been made, and hence we place a stringent upper limit on the abundance of PBHs at $95\%$ confidence level. In particular, less than one part in a million of the total DM mass could come from PBHs in the mass range of $[2 \times 10^{-3}, 7\times 10^{-1}] \Msun$.

\end{abstract}
	
\pacs{???}
	
\maketitle

{\it Introduction. }
Over the past few years, the great achievement of detecting gravitational waves (GWs)
from binary black holes (BBHs) \cite{Abbott:2016blz,Abbott:2016nmj,Abbott:2017vtc,%
Abbott:2017gyy,Abbott:2017oio,TheLIGOScientific:2016pea,LIGOScientific:2018mvr} 
and a binary neutron star (BNS) \cite{TheLIGOScientific:2017qsa} by \lvc\ 
has led us to the era of GW astronomy, 
as well as the era of multi-messenger astronomy.
Various models 
have been proposed to account for the formation 
and evolution of these \lvc\ BBHs, 
among which the PBH scenario \cite{Bird:2016dcv,Sasaki:2016jop,Chen:2018czv}
has attracted a lot of attention recently.
PBHs are predicted to undergo gravitational collapse 
from overdensed regions in the infant universe \cite{Hawking:1971ei,Carr:1974nx} when the corresponding 
wavelength of enhanced scalar curvature perturbations re-enter the horizon 
\cite{Ivanov:1994pa,Yokoyama:1995ex,GarciaBellido:1996qt,Ivanov:1997ia,Kawasaki:2006zv}. 

The PBH scenario is appealing because it can not only account for the event 
rate of \lvc\ BBHs, but also be a promising candidate for the long elusive
missing part of our Universe -- dark matter (DM). 
It is inconclusive that whether PBH can represent all DM or not, 
yet the abundance of PBHs ($\fpbh$) which describes the total DM mass in the form of PBHs, 
has been constrained by a variety of observations, 
such as extra-galactic $\gamma$-rays from PBH evaporation \cite{Carr:2009jm}, femtolensing of $\gamma$-ray bursts \cite{Barnacka:2012bm}, Subaru/HSC microlensing \cite{Niikura:2017zjd}, 
Kepler milli/microlensing \cite{Griest:2013esa}, 
OGLE microlensing \cite{Niikura:2019kqi},
EROS/MACHO microlensing \cite{Tisserand:2006zx},
existence of white dwarfs (WDs) which are not triggered to explode in our local galaxy \cite{Graham:2015apa} (this constraint might be ineffective according to the simulation in \cite{Montero-Camacho:2019jte}),
dynamical heating of ultra-faint dwarf galaxies \cite{Brandt:2016aco}, 
X-ray/radio emission from the accretion of interstellar gas onto PBHs \cite{Gaggero:2016dpq}, cosmic microwave background radiation from the accretion of primordial gas onto PBHs \cite{Ali-Haimoud:2016mbv,Blum:2016cjs,Horowitz:2016lib,Chen:2016pud},
and GWs either through the null detection of sub-solar mass BBHs 
\cite{Abbott:2018oah,Magee:2018opb,Chen:2019irf,Authors:2019qbw}
or the null detection of stochastic GW background (SGWB) from BBHs \cite{Wang:2016ana,Chen:2019irf}. But PBHs in a substantial window in the approximate mass range $[10^{-16},10^{-14}] \cup [10^{-13},10^{-12}] M_\odot$ are still allowed to account for all of the DM. We refer to \cite{Chen:2019irf} for a recent summary.

Actually there is another way to probe the PBH DM scenario, 
namely through the scalar induced GWs (SIGWs) which would inevitably be generated in conjunction with the formation of PBHs \cite{tomita1967non,Saito:2008jc,Young:2014ana,Cai:2018dig,Yuan:2019udt,Cai:2019elf,Yuan:2019fwv}.
The feature for distinguishing SIGW from other sources was sketched out in \cite{Yuan:2019wwo} recently.
Since PBHs are supposed to form from the tail of the probability density 
function of the curvature perturbations, 
the possibility to form a single PBH is quite sensitive to the amplitude 
of curvature perturbation power spectrum \cite{Young:2014ana}.
Consequently the abundance of PBHs is extremely sensitive to the amplitude of the corresponding SIGW.
Therefore a detection of SIGW will provide evidence for PBHs, 
while the null detection of SIGW will put a stringent constraint on the abundance of PBHs. 

The peak frequency of the SIGW $(f_*)$ is determined by the peak wave-mode of 
the comoving curvature power spectrum, 
and thus is related to the mass of PBHs by $f_*\sim 3\,{\rm{Hz}}\({m_{\rm{pbh}}/10^{-18} M_\odot}\)^{-1/2}$ \cite{Saito:2008jc}.
The mass of PBHs constituting DM should be heavier than $10^{-18}\Msun$, 
otherwise they would have evaporated due to Hawking radiation.
As a result, the corresponding peak frequency of the SIGW should be lower than $3$Hz, and then it is difficult for the ground-based detectors like \lvc\ to detect the corresponding SIGWs. 
On the other hand, the GW observatories hunting for low frequency signals are especially
suitable to explore the PBH DM hypothesis, 
and the prospective constraints on the abundance of PBHs by LISA \cite{Audley:2017drz} 
and pulsar timing observations such as IPTA \cite{Hobbs:2009yy}, FAST \cite{Nan:2011um} 
and SKA \cite{Kramer:2015jsa} have been investigated in \cite{Yuan:2019udt}. See some other related works in \cite{Inomata:2016rbd,Schutz:2016khr,Orlofsky:2016vbd,Dror:2019twh,Wang:2019kaf,Cai:2019elf,Clesse:2018ogk}.

Despite the data of current pulsar timing array (PTA) has been used to constrain the amplitude of SGWBs, those results strongly depend on the assumption of some special power-law form which is quite different from SIGWs \cite{Yuan:2019udt}. Therefore, in this article, we perform the first search in the public available PTA data set for the signal of SIGWs in order to test the PBH DM hypothesis. In particular, the null detection of SIGWs in the current NANOGrav 11-year data set \cite{Arzoumanian:2017puf} provides a constraint on the abundance of PBHs through SIGWs in the mass range of $[4\times 10^{-4}, 1.7]\Msun$.


{\it PBH DM and SIGW.}
In this article, we consider the monochromatic formation of PBHs, corresponding to a $\delta$ power spectrum of the scalar curvature perturbation, i.e. 
\e
\mathcal{P}_{\zeta}(f)=A f_*\delta\(f-f_*\),
\q
where $A$ is the dimensionless amplitude of the power spectrum. In this case, the mass of the PBHs is related to the peak frequency $f_*$ by, \cite{Hawking:1971ei,Carr:1974nx}, 
\m\label{mkrelation}
{m_{\mathrm{pbh}}\over M_{\odot}} \simeq 2.3\times10^{18}
\left(\frac{H_{0}}{f_*}\right)^{2},
\n
where $f_*$ is in units of Hz, and $H_0$ is the Hubble constant. 
The formation of PBH is a threshold process which is described by three-dimensional statistics of Gaussian random fields, also known as peak theory \cite{Bardeen:1985tr}, and the abundance of PBH in DM,
$f_{\mathrm{pbh}}\equiv\Omega_{\mathrm{pbh}}/\Omega_{\mathrm{DM}}$, is given by, \cite{Carr:2016drx}, 
\m\label{fpbh}
f_{\mathrm{pbh}} \simeq 1.9 \times 10^{7}
  \({\zeta_c^2/A}-1\) e^{-{\zeta_c^2\over 2A}} \left(\frac{m_{\mathrm{pbh}}}{M_{\odot}}\right)^{-\frac{1}{2}},
\n
where $\zeta_{c}\simeq1$ \cite{Musco:2008hv,Musco:2004ak,Musco:2012au,Harada:2013epa,Escriva:2019nsa,Escriva:2019phb} is the threshold value for the formation of PBHs.

In \cite{Maggiore:1999vm} the energy density of a GW background $\rho_{\mathrm{GW}}$ takes the form  
\e
\rho_{\mathrm{GW}} = \int\rho_{\mathrm{GW}}(f,\eta)\ \rd\ln f ={M_p^2\over16a^2}\<\overline{\partial_kh_{ij}\partial^kh^{ij}}\>,
\q
where $\eta$ is the conformal time, $a$ is the scale factor, $M_p$ is the Planck mass, and the overline stands for time average. It is useful to introduce the dimensionless GW energy density parameter per logarithm frequency $\ogw(\eta, k)$ defined by
\e
\ogw(\eta, f)\equiv\frac{\rho_{\mathrm{GW}}(f,\eta)}{\rho_{\rm{cr}}},
\q
where $\rho_{\rm{cr}}$ is the critical energy of the present Universe. 
For a monochromatic formation of PBHs, the present $\Omega_{\mathrm{GW}}(f)$ of the SIGW in radiation dominated era can be estimated as \cite{Yuan:2019udt}
\e\label{ogw}
\ogw(f) = \ogw^{(2)}(f) + \ogw^{(3)}(f).
\q 
Here, the leading order contribution $\ogw^{(2)}(f)$ is given by, \cite{Espinosa:2018eve,Kohri:2018awv}, 
\m
\ogw^{(2)}(f)&&={3\tilde{f}^2A^2\over 1024}\Omega_r(4-\tilde{f}^2)^2(3\tilde{f}^ 2-2)^2\, \Theta(2-\tilde{f}) \times\no\\
&&\qquad\Big[\pi^2(3\tilde{f}^2-2)^2\, \Theta(2\sqrt{3}-3\tilde{f})\no\\
&&\qquad\quad+\(4+(3\tilde{f}^2-2)\log|1-\frac{4}{3\tilde{f}^2}|\)^2\Big],
\n
where $\tilde{f}\equiv f/f_*$ is the dimensionless frequency 
and $\Theta$ is the Heaviside theta function. 
In addition, the third-order correction $\ogw^{(3)}(f)$ reads, \cite{Yuan:2019udt}, 
\e
\ogw^{(3)}(f)=\frac{A^3}{384\tilde{f}^{2}} \Omega_r \(M_2\overline{I_3^2}+M_1\overline{I_2 I_4}\).
\q
The definitions of $M_1$, $M_2$, $I_2$, $I_3$, and $I_4$ are complicated and
can be found in \cite{Yuan:2019udt}.

{\it PTA data analysis. }
Null detection of certain GW backgrounds has been reported by the current PTAs
such as NANOGrav\footnote{\url{http://nanograv.org}}, 
PPTA\footnote{\url{https://www.atnf.csiro.au/research/pulsar/ppta}} 
and EPTA\footnote{\url{http://www.epta.eu.org}},
and the upper bounds on the amplitude of those GW backgrounds have also been continuing improved.
For instance, NANOGrav constrained on the SGWB produced by supermassive
black holes \cite{Aggarwal:2018mgp} and other spectra \cite{Arzoumanian:2018saf}
such as power-law, broken-power-law, free and Gaussian-process ones. 
Similar studies were also performed by the PPTA  collaboration \cite{Shannon:2013wma} and the EPTA collaboration \cite{vanHaasteren:2011ni}.
In this article we search for the signal of SIGW using the NANOGrav 11-year data set 
which consists of time of arrival (TOA) data and pulsar timing models presented in \cite{Arzoumanian:2017puf}.
Similar to \cite{Kato:2019bqz}, we choose six pulsars  
which have relatively good TOA precision and long observation time. 
A summary of the basic properties of these pulsars is presented 
in \Table{pulsars}.
For all the $6$ pulsars, $T_{\rm{obs}}$ is longer than $8$ years, $N_{\rm{TOA}}$ is more than $10^4$, and RMS is less than $1.5\mu s$. 
\begin{table}[htb]
    \caption{\label{pulsars} Basic properties of the $6$ pulsars used in 
        our analysis: RMS - the weighted root-mean-square epoch-averaged post-fit timing residuals,
        $N_{\rm{epoch}}$ - number of observational epochs,
        $N_{\rm{TOA}}$ - number of TOAs,
        $T_{\rm{obs}}$ - observational time span.
        See Ref.~\cite{Arzoumanian:2017puf} in detail.}        
    \begin{tabular}{ccccc}
        \hline\hline
        Pulsar name\hspace{1mm} & RMS [$\mu$s]\hspace{1mm} & $N_{\rm{epoch}}$\hspace{1mm} & $N_{\rm{TOA}}$\hspace{1mm} & $T_{\rm{obs}}$ [yr] \\
        \hline    
        J0613$-$0200 & 0.422 & 324 & 11,566 & 10.8  \\
        J1012$+$5307 & 1.07 & 493 & 16,782 & 11.4  \\
        J1600$-$3053 & 0.23 & 275 & 12,433 & 8.1  \\
        J1713$+$0747 & 0.108 & 789 & 27,571 & 10.9  \\
        J1744$-$1134 & 0.842 & 322 & 11,550 & 11.4  \\
        J1909$-$3744 & 0.148 & 451 & 17,373 & 11.2  \\    
        \hline \hline
    \end{tabular}
\end{table}


The presence of a GW background will manifest as the unexplained residuals in the TOAs of pulsar signals after
subtracting a deterministic timing model that accounts for the pulsar spin behavior and the geometric effects due to the 
motion of the pulsar and the Earth \cite{1978SvA....22...36S,Detweiler:1979wn}.
It is therefore feasible to separate GW-induced residuals, which have
distinctive correlations among different pulsars \cite{Hellings:1983fr}, 
from other systematic effects, such as clock errors or delays due to light propagation through interstellar medium,
by regularly monitoring TOAs of pulsars from an array of the most rotational stable millisecond pulsars 
\cite{1990ApJ...361..300F}.
An $\NTOA$ length vector $\dt \bm{t}$ representing the timing residuals 
for a single pulsar can be modeled as follows \cite{Taylor:2012wv,vanHaasteren:2012hj}
\e\label{dt}
\dt\bm{t} = \bm{M} \bm{\epsilon} + \bn{RGP},
\q
where $\bm{M}$ is the timing model design matrix, $\bm{\epsilon}$ is a
vector denoting small offsets for the timing model parameters,
and $\bm{M} \bm{\epsilon}$ is the residual due to inaccuracies 
of the timing model.
The timing model design matrix is obtained through  \texttt{libstempo}\footnote{\url{https://vallis.github.io/libstempo}} 
package which is a python interface to 
\texttt{TEMPO2} \footnote{\url{https://bitbucket.org/psrsoft/tempo2.git}}
\cite{Hobbs:2006cd,Edwards:2006zg} timing software.
The term $\bn{RGP}$ in \Eq{dt} is the stochastic contribution to the TOAs, which can be modeled by a sum of random Gaussian processes \cite{vanHaasteren:2014qva} as
\e\label{noise}
\bn{RGP} = \bn{RN} + \bn{WN} + \bn{SSE} + \bn{SIGW}.
\q

The first term on the right hand side of \Eq{noise}, $\bn{RN}$, 
represents the red noise via a Fourier decomposition,
\e 
\bn{RN} = \sum_{j=1}^\Nmode \[a_j \sin\(\frac{2\pi j t}{T}\)
		+ b_j \cos\(\frac{2\pi j t}{T}\)\] 
		= \bm{F} \bm{a},
\q 
where $\Nmode$ is the number of frequency modes included in the sum,
$T$ is the total observation time span, 
$\bm{F}$ is the Fourier design matrix with components of alternating
sine and cosine functions for frequencies in the range $[1/T, \Nmode/T]$,
and $\bm{a}$ is a vector giving the amplitude of the Fourier basis functions.
In the analysis, we choose $\Nmode=50$.
The covariant matrix of the red noise coefficients $\bm{a}$ at frequency
modes $i$ and $j$ will be diagonal, namely
\e\label{aa}
\av{\bm{a}_i \bm{a}_j} = P(f_i)\, \dt_{ij},
\q 
where the power spectrum $P(f)$ is usually well described by a power-law model,
\e 
P(f) = \frac{\ARN^2}{12\pi^2} \(\frac{f}{\rm{yr}^{-1}}\)^{3-\gRN} f^{-3},
\q 
with $\ARN$ and $\gRN$ the amplitude and spectral index of the power-law, 
respectively.
Note that in \Eq{aa}, $f_i$ is defined by $i/T$ if $i$ is odd, 
and $(i-1)/T$ if $i$ is even.

The second term, $\bn{WN}$, accounts for the influence of white noise on
the timing residuals, including a scale parameter on the TOA uncertainties
(EFAC), an added variance (EQUAD) and a per-epoch variance (ECORR) for 
each backend/receiver system.
This white noise is assumed to follow Gaussian distribution and can
be characterized by a covariance matrix as
\e 
\bC{WN} = \bC{EFAC} + \bC{EQUAD} + \bC{ECORR}, 
\q 
where $\bC{EFAC}$, $\bC{EQUAD}$ and $\bC{ECORR}$ are the correlation functions
for EFAC, EQUAD and ECORR parameters, respectively.
Explicit expressions for these correlation functions can be found in \cite{Kato:2019bqz}. 

The third term, $\bn{SSE}$, is a noise due to inaccuracies of a solar system
ephemeris (SSE) which is used to convert observatory TOAs to an inertial frame
centered at the solar system barycenter.
The SSE noise can seriously affect the upper limits and Bayes factors when
searching for stochastic gravitational-wave backgrounds \cite{Arzoumanian:2018saf}.
In our analysis, we use DE436 \cite{DE436} as the fiducial SSE model. 
To account for the SSE errors, we employ the physical model \textsc{BayesEphem} 
introduced in \cite{Arzoumanian:2018saf} and implemented in NANOGrav's 
flagship package 
\texttt{enterprise}\footnote{\url{https://github.com/nanograv/enterprise}}.
The \textsc{BayesEphem} model has eleven parameters, including four parameters
correspond to perturbations in the masses of the outer planets,
one parameter describes a rotation rate about the ecliptic pole,
and six parameters characterize the corrections to Earth's orbit generated by
perturbing Jupiter's average orbital elements \cite{Arzoumanian:2018saf}.

The last term, $\bn{SIGW}$, is the observed timing residuals due to the SIGW,
which are described by the cross-power
spectral density \cite{Thrane:2013oya} 
\e
S_{IJ}(f)=\frac{H_0^2}{16\pi^4f^5}\Gamma_{IJ}(f)\ \Omega_{\mathrm{GW}}(f),
\q
where $\Gamma_{IJ}$ is the Hellings \& Downs coefficients \cite{Hellings:1983fr}
measuring the spatial correlation of the pulsars $I$ and $J$ in the array.
The expression for $\ogw(f)$ is given by \Eq{ogw}.
The free parameters for the SIGW are the amplitude $A$ and the peak frequency $f_*$.
For a fixed $f_*$, the mass of PBH is given by Eq.~(\ref{mkrelation}). 
In this sense, the free parameter $A$ is directly related to the abundance of
PBHs $f_{\rm{pbh}}$. 

\begin{table*}[htb!]
\scriptsize
\caption{Parameters and their prior distributions used in the analyses.}
\label{tab:priors}
\begin{tabular}{llll}
\hline\hline
parameter & description & prior & comments \\
\hline
\multicolumn{4}{c}{SIGW signal} \\[1pt]
$A$ & GWB strain amplitude & Uniform $[10^{-5}, 10^{0}]$ (upper limits) & \\
& & log-Uniform $[-5, 0]$ (model comparison) & one parameter for PTA \\
$\fstar$ & peak frequency & delta function  & fixed \\
\hline
\multicolumn{4}{c}{White Noise} \\[1pt]
$E_{k}$ & EFAC per backend/receiver system & Uniform $[0, 10]$ & single-pulsar analysis only \\
$Q_{k}$[s] & EQUAD per backend/receiver system & log-Uniform $[-8.5, -5]$ & single-pulsar analysis only \\
$J_{k}$[s] & ECORR per backend/receiver system & log-Uniform $[-8.5, -5]$ & single-pulsar analysis only \\
\hline
\multicolumn{4}{c}{Red Noise} \\[1pt]
$\ARN$ & red-noise power-law amplitude & Uniform $[10^{-20}, 10^{-11}]$ (upper limits) & \\
& & log-Uniform $[-20, -11]$ (model comparison) & one parameter per pulsar  \\
$\gRN$ & red-noise power-law spectral index & Uniform $[0, 9]$ & one parameter per pulsar \\
\hline
\multicolumn{4}{c}{\textsc{BayesEphem}} \\[1pt]
$z_{\rm drift}$ [rad/yr] & drift-rate of Earth's orbit about ecliptic $z$-axis & Uniform [$-10^{-9}, 10^{-9}$] & one parameter for PTA \\
$\Delta M_{\rm jupiter}$ [$M_{\odot}$] & perturbation to Jupiter's mass & $\mathcal{N}(0, 1.55\times 10^{-11})$  & one parameter for PTA \\
$\Delta M_{\rm saturn}$ [$M_{\odot}$] & perturbation to Saturn's mass & $\mathcal{N}(0, 8.17\times 10^{-12})$  & one parameter for PTA \\
$\Delta M_{\rm uranus}$ [$M_{\odot}$] & perturbation to Uranus' mass & $\mathcal{N}(0, 5.72\times 10^{-11})$  & one parameter for PTA \\
$\Delta M_{\rm neptune}$ [$M_{\odot}$] & perturbation to Neptune's mass & $\mathcal{N}(0, 7.96\times 10^{-11})$  & one parameter for PTA \\
PCA$_{i}$ & principal components of Jupiter's orbit & Uniform $[-0.05, 0.05]$ & six parameters for PTA \\
\hline
\end{tabular}
\end{table*}

For the timing model parameters and TOAs, we use the publicly available data files from NANOGrav 11-year data set \cite{Arzoumanian:2017puf}.
To extract information from the data, we perform a Bayesian inference 
by closely following the procedure in \cite{Arzoumanian:2018saf}. 
The parameters of our model and their prior distributions are presented in 
\Table{tab:priors}.
In order to reduce the computational costs, a common strategy is to fix the
white noise parameters to their max likelihood values determined from 
independent single-pulsar analysis, 
in which only the white and red noises are considered.
Fixing white noise parameters can greatly reduce the number of free parameters.

Assuming the $\bn{RGP}$ is Gaussian and stationary, 
for a PTA with M pulsars, the 
likelihood function can be evaluated as, \cite{Ellis:2013nrb},
\e
\mathcal{L} = \frac{1}{\sqrt{\det(2\pi \bS)}} \exp\(-\hf \bN^T \bS^{-1} \bN\),
\q 
where $\bN \equiv \[\bn{RGP}^1, \bn{RGP}^2, \cdots, \bn{RGP}^M\]^T$
is a collection of $\bn{RGP}$ for all pulsars,
and $\bS \equiv \av{\bN \bN^T}$ is the covariance matrix. 
Following the common practice in \cite{Lentati:2012xb,vanHaasteren:2014qva,%
vanHaasteren:2014faa}, 
we marginalize over the timing model parameter $\bm{\epsilon}$ when evaluating
the likelihood.
The likelihood is calculated by using the pulsar timing package
\texttt{enterprise}.
To achieve parallel tempering, we use \texttt{PTMCMCSampler}\footnote{\url{https://github.com/jellis18/PTMCMCSampler}}
package to do the Markov chain Monte Carlo sampling. 

Given the observational data $\mathcal{D}$, 
one needs to distinguish two exclusive models: 
a noise-only model $\mathcal{H}_0$ and a noise-plus-signal model $\mathcal{H}_1$.
The model selection is quantified by the Bayes factor
\e\label{bayes} 
    B_{10} = \frac{\rm{evidence}[\mathcal{H}_1]}{\rm{evidence}[\mathcal{H}_0]}
        = \frac{p(A=0|\mathcal{H}_1)}{p(A=0|\mathcal{D},\mathcal{H}_1)},
\q 
where the numerator and denominator are the prior and posterior probability 
density of $A=0$ in the model $\mathcal{H}_1$, respectively.
We have used the Savage-Dickey formula \cite{dickey1971weighted} 
to estimate the Bayes factor in \Eq{bayes}.

\begin{figure}[htbp!]
	\centering
	\includegraphics[width = 0.5\textwidth]{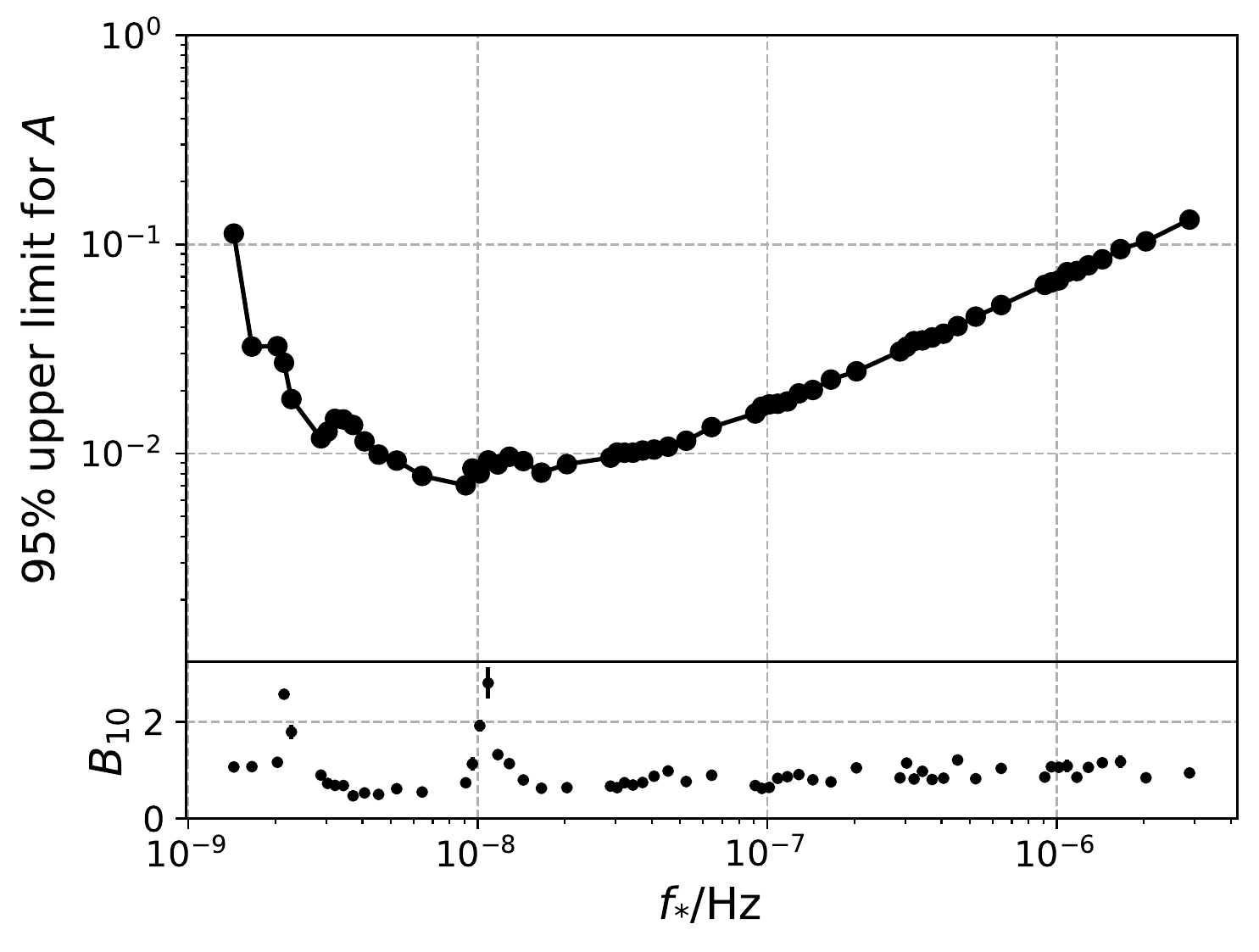}
	\caption{\label{A_upper} \textbf{Top panel}: the $95\%$ upper limits on the
        power spectrum amplitude $A$ of curvature perturbation as a function
        of the peak frequency $\fstar$ from the NANOGrav 11-year data set.
        \textbf{Bottom panel}: the corresponding Bayes factors $B_{10}$ as a
        function of the peak frequency $\fstar$.
	}
\end{figure}

\begin{figure}[htbp!]
\centering
\includegraphics[width = 0.5\textwidth]{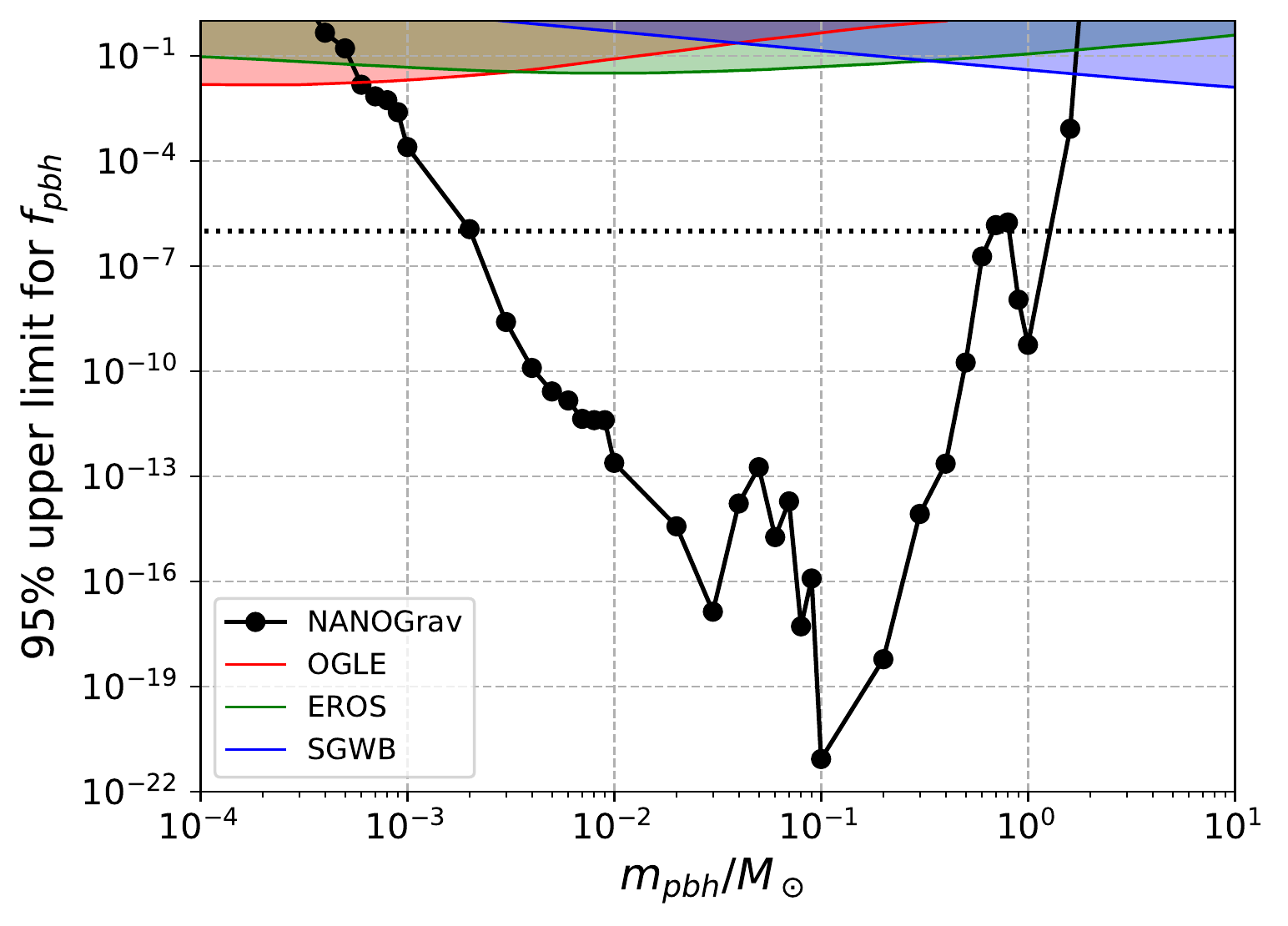}
\caption{\label{fpbh_upper} The $95\%$ upper limits on the abundance of PBHs
    in DM $\fpbh$ as a function of the PBH mass $m_{\rm{pbh}}$ from the
    NANOGrav 11-year data set. 
    Results from OGLE microlensing (OGLE) \cite{Niikura:2019kqi},
    EROS/MACHO microlensing (EROS) \cite{Tisserand:2006zx}, 
    and SGWB \cite{Chen:2019irf} are also shown. 
    The horizontal dotted line corresponds to $10^{-6}$.}
\end{figure}

{\it Results and conclusion.  } The upper limits and the Bayes factor
for the power spectrum amplitude $A$ as a function of the peak frequency $\fstar$ from the NANOGrav 11-year data set are showed in \Fig{A_upper} at the $95\%$ confidence level. 
Even though there are two peaks in the Bayes factor distribution, both peak values are smaller than $3$, implying the presence of a signal in the data is ``not worth more than a bare mention" \cite{BF}.
Since the Bayes factor $B_{10}$ for each peak frequency is less than $3$, it indicates that the data is consistent with containing noise only. 
The upper limits on the abundance of PBHs in DM $\fpbh$ as a function of the PBH mass $m_{\rm{pbh}}$ are given in \Fig{fpbh_upper} at the $95\%$ confidence level. 
Note that $m_{\rm{pbh}}$ is related to $\fstar$ by \Eq{mkrelation}, 
and $\fpbh$ is related to $A$ and $m_{\rm{pbh}}$ by Eq.~\eqref{fpbh}. 
Our results imply that the current PTA data set has already been able to place a stringent
constraint on the abundance of PBHs through the SIGWs. 
According to \Fig{fpbh_upper}, the abundance of PBHs is less than $10^{-6}$ in the mass range of $[2 \times 10^{-3}, 7\times 10^{-1}] M_\odot$.

In this article, we give the first search for the signal of SIGWs inevitably accompanying the formation of PBHs in the NANOGrav 11-year data set.
Since no significant signal is found, we place a $95\%$ upper limit on the amplitude of scalar perturbation over the peak frequency range of $[1.5\times 10^{-9}, 3\times 10^{-6}]$Hz and the 
abundance of PBHs in the mass range of $[4\times10^{-4}, 1.7]\Msun$. 
In particular, the abundance of PBHs in the mass range of $[2 \times 10^{-3}, 7\times 10^{-1}] M_\odot$ less than $10^{-6}$, which is much better than any other observational constraints in this mass range in literature.
Since the amplitude of SIGWs is roughly determined by the peak amplitude of scalar power spectrum even for the case with an extended mass distribution, a similar constraint on the peak amplitude of scalar power spectrum should be obtained from NANOGrav 11-yr data, and therefore a stringent constraint on the abundance of PBHs with an extended mass distribution can be also expected. In principle, the exact analysis for the case with an extended mass distribution is model-dependent, and will be left for the future.



{\it Acknowledgments. }
We acknowledge the use of HPC Cluster of ITP-CAS. 
This work is supported by grants from NSFC 
(grant No. 11975019, 11690021, 11991052, 11947302), 
the Strategic Priority Research Program of Chinese Academy of Sciences 
(Grant No. XDB23000000, XDA15020701), and Key Research Program of Frontier Sciences, CAS, Grant NO. ZDBS-LY-7009.
	
\bibliography{./ref}

\end{document}